\documentclass[11pt]{article}
\usepackage[utf8]{inputenc}
\usepackage{graphicx}
\usepackage{amsmath}
\usepackage{url}
\usepackage{hyperref}
\usepackage[margin=1in]{geometry}
\usepackage{authblk}

\title{Automatic Identification of Magnetospheric Regions using Supervised Machine Learning Models}

\author[1]{Narges Ahmadi}
\author[1]{Robert E. Ergun}
\author[1]{Xiangning Chu}
\author[1]{Victoria Wilder}

\affil[1]{Laboratory of Atmospheric and Space Physics, University of Colorado Boulder, Boulder, Colorado 80303, USA}

\date{}

\begin{document}

\maketitle

\begin{abstract}

\bigskip

We present an automated approach for identifying magnetospheric regions using supervised machine learning techniques applied to Magnetospheric MultiScale mission data. Our method utilizes ion energy spectra, total magnetic field, total ion temperature, ion velocity component, ion density and spacecraft position data to classify five distinct plasma environments: solar wind, magnetosheath, inner magnetosphere, plasma sheet, and lobe regions. The approach combines a convolutional neural network (CNN) for analyzing ion energy spectrogram data with a Random Forest classifier for scalar plasma parameters. The CNN method employs 2D convolution to identify spatial and temporal patterns in the ion energy spectrogram treated as image-like data, while the Random Forest model processes averaged magnetic field, temperature, velocity, density and position parameters. Our hybrid model achieves 99\% accuracy on test dataset with an F1 score of 0.99, providing reliable automated region identification at 3-minute temporal resolution. This lightweight approach requires minimal manual data labeling and can be readily applied to other magnetospheric missions with similar data products.
\end{abstract}

\bigskip

\textbf{Key Points:}
\begin{itemize}
\item Ion energy spectra, magnetic field, ion moments, and position 
data enable automated identification of five magnetospheric regions.
\item Combined Convolutional Neural Network and Random Forest processes both spectrogram and scalar data for comprehensive pattern recognition.
\item The model achieves 99\% accuracy while maintaining computational efficiency suitable for operational use.
\end{itemize}

\section*{Plain Language Summary}
Identifying magnetospheric regions in space plasmas around Earth helps us to study the dynamics of each region and how it impacts Earth's space weather. We developed a computer program that can automatically identify which region of space around Earth a spacecraft is located in, eliminating the need for time-consuming manual analysis. The program analyzes spacecraft measurements like particle energy levels, magnetic field strength, ion temperature, density, velocity and position data to distinguish between five different space environments, including areas like the solar wind and inner magnetosphere. It uses two different methods working together: one treats particle energy data like images to spot patterns, while another processes instrument measurements like magnetic field and temperature. The automated system achieves 99\% accuracy and can make identifications every 3 minutes without requiring much computer power or extensive manual training data, making it a valuable time saving tool that can be easily adapted for other space missions.

\section{Introduction}

Space plasma missions generate vast quantities of data requiring systematic analysis to understand magnetospheric dynamics and space weather phenomena. Accurate identification of distinct plasma environments including the magnetosphere, magnetosheath, solar wind, plasma sheet, and lobe regions represents a fundamental prerequisite for scientific analysis. Solar wind is a continuous stream of charged particles flowing outward from the Sun at speeds of 300-800 km/s. It carries the interplanetary magnetic field and provides the energy that drives magnetospheric dynamics when it interacts with planetary magnetic fields. Magnetosheath is the turbulent region between the bow shock and magnetopause where the solar wind has been slowed, heated, and compressed after passing through the bow shock. The plasma here is denser and hotter than the original solar wind but still flows around the magnetosphere. Inner magnetosphere is the region close to Earth extending from the upper atmosphere to about 6-10 Earth radii, dominated by Earth's dipolar magnetic field. Plasma Sheet is a thin, current carrying layer of hot plasma located in the center of the magnetotail, typically extending from about 10 to 100+ Earth radii downstream. It's a key site for magnetic reconnection and particle acceleration during substorms. Lobe is the region above and below the plasma sheet in the magnetotail, characterized by low temperature and magnetic field lines that are stretched tailward. Experienced researchers can visually distinguish these regions through inspection of plasma parameters and particle energy spectrogram data but the task can be labor intensive for large datasets and statistical studies. 

Multiple large-scale magnetospheric missions currently orbit Earth, collecting substantial volumes of observational data. These include the Magnetospheric Multiscale (MMS) mission launched in 2015 \cite{Burch2015}, the Time History of Events and Macroscale Interactions during Substorms (THEMIS) mission launched in 2007 \cite{Angelopoulos2008}, and the Cluster mission operational from 2000 to 2024 \cite{Escoubet2001}. Scientific analysis of these datasets demands accurate plasma environment classification as a foundational step in data processing workflows. These missions also face data downlink constraints due to limited deep space network contact time and onboard storage capacity for high resolution data. For instance, the MMS mission downlinks only approximately 4\% of its recorded high-resolution data. Currently, the mission relies on Scientist-in-the-Loop (SITL), a system operational since the 2015 launch in which trained scientists manually scan low resolution data to select intervals for high resolution downlink. This labor intensive process requires a volunteer scientist weekly. This work aims to automate data selection for MMS and future magnetospheric missions, eliminating this operational bottleneck.

Traditional region identification approaches rely on visual inspection and empirically determined thresholds applied to plasma parameters. For dayside regions, several studies have established threshold-based classification methods \cite{Karlsson2021, Raptis2020, Jelinek2012}. Magnetotail region identification typically employs threshold conditions combined with statistical analysis of background plasma parameters using Cluster and MMS observations \cite{Boakes2014, Vo2023}. However, these approaches require significant manual oversight and may not generalize effectively across different space weather conditions or mission configurations.

Recent developments in machine learning have enabled automated classification approaches for magnetospheric regions. Breuillard et al. (2020) developed a fully convolutional network (FCN) achieving 89\% accuracy in identifying 10 magnetospheric regions using 12 plasma parameters at 3-minute resolution from MMS data. Argall et al. (2020) employed Long Short-Term Memory networks with 123 parameters for magnetopause crossing detection at 4.5-second resolution. Olshevsky et al. (2021) utilized 3D particle energy distributions for dayside region classification, achieving greater than 98\% accuracy. Additional studies have applied machine learning to specific boundary detections \cite{Lalti2022} and employed unsupervised approaches for region classification \cite{Toy-Edens2024, Waters2024}. Wang et al. (2025) introduced a wavelet-decision tree classifier designed to automate dayside region detection. Nguyen et al. (2022) developed a lightweight gradient boosting classifier using magnetic field and plasma moment data to identify three magnetospheric regions that works for multiple missions. However, current automated classification methodologies lack the capability to systematically process large datasets with both comprehensive regional coverage and high accuracy.

This study presents a lightweight, high-accuracy supervised learning model specifically designed to identify all primary magnetospheric regions using MMS data. Our approach combines complementary machine learning techniques to process different data types: a convolutional neural network for ion energy spectrograms and a Random Forest classifier for scalar plasma parameters. This hybrid methodology achieves superior performance while maintaining computational efficiency and minimal training data requirements.

\section{Data and Methodology}

\subsection{The MMS Mission}

The MMS mission employs four identical spacecraft in a controlled tetrahedral formation to study magnetic reconnection processes throughout Earth's magnetosphere \cite{Burch2015}. The mission's highly elliptical orbit maximizes observational time in regions where magnetic reconnection occurs, providing comprehensive coverage of near-Earth plasma environments.
We only use MMS1 for our data collection since all spacecraft are identical. Our analysis utilizes data from multiple MMS1 instrument suites. Ion energy spectra measurement, ion density, ion velocity and ion temperature are provided by the Fast Plasma Investigation (FPI) instrument \cite{Pollock2016} at 4.5 second temporal resolution in survey mode. Magnetic field data are obtained from the Fluxgate Magnetometer (FGM) within the FIELDS instrument suite \cite{Russell2016, Torbert2016} at 0.1 second resolution. Spacecraft position information is derived from the Magnetic Ephemeris and Coordinates (MEC) data product \cite{Henderson2022}.

For this study, we construct 3-minute data segments (40 data points at 4.5 second resolution) from four primary measurements: ion energy spectrograms, total magnetic field, total ion temperature, ion density, ion velocity component in Geocentric Solar Ecliptic (GSE) X-coordinate and also spacecraft position in GSE X-coordinate. Each 3-minute segment receives a single region label, providing 3-minute temporal resolution for region identification.

\subsection{Training Data Generation and Labeling}

\begin{figure}[!ht]
\centering
\includegraphics[width=.98\textwidth]{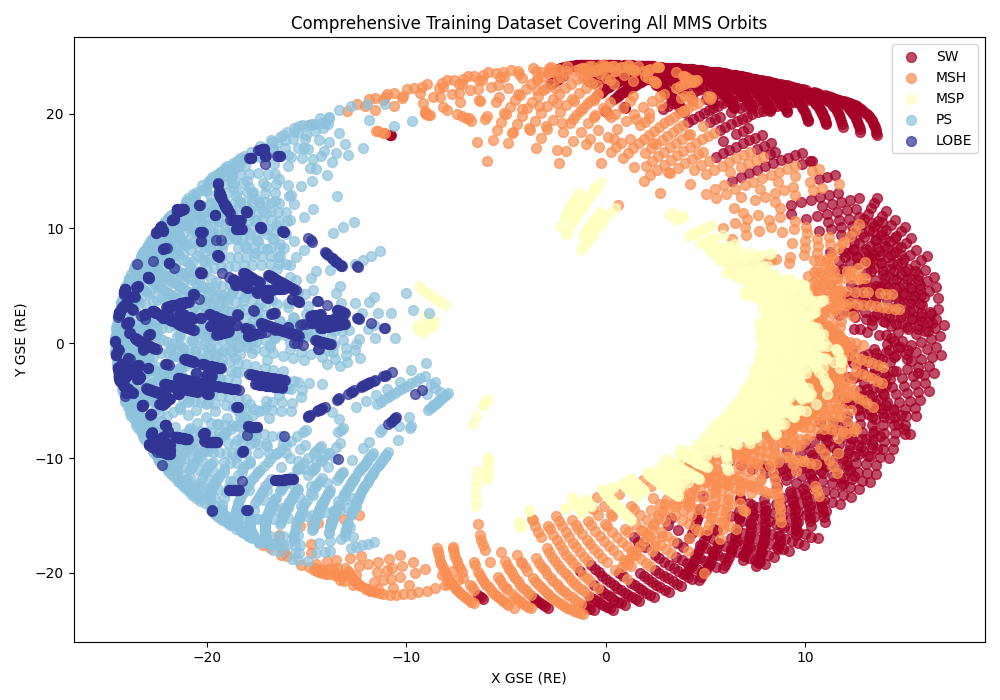}
\caption{\label{fig:AIdata} Magnetospheric regions labeled in one year of MMS data. Red: solar wind, orange: magnetosheath, yellow: inner magnetosphere, light blue: plasma sheet, dark blue: lobe}
\end{figure}

We identify five primary magnetospheric regions: solar wind (SW), magnetosheath (MSH), inner magnetosphere (MSP), plasma sheet (PS), and lobe (LOBE). 
We have prepared a training/validation/test dataset for the ML model using one year (year 2017) of publicly available MMS data for a comprehensive coverage of all orbits. We only use MMS1 for our data collection since all spacecraft are identical at the temporal resolution we are using. Figure \ref{fig:AIdata} demonstrates the training dataset and the associated labels. The dataset has undergone manual verification by experienced space plasma researchers. Following quality control, representative samples per region class are selected, yielding a closely balanced training set of 8979 samples (Table~\ref{table:1}).

 The number of different labels can be highly imbalanced, with some regions overwhelmingly more frequent than others. Data imbalance is a common issue in space physics \cite{Camporeale2019} and has been addressed in many recent studies \cite{Chu2023, Chu2024, Chu2025}. Our representative samples for each region class is relatively balanced as shown in table \ref{table:1}. The dataset was partitioned into training (60\%), validation (20\%), and test (20\%) subsets using stratified sampling to ensure each subset maintained the same class label distribution as the original dataset. This partitioning ensures robust model evaluation while providing sufficient training data for effective learning. 

\begin{table}[h!]
\centering
\begin{tabular}{ c c c c c c}
\hline
Region & SW & MSH & MSP & PS & LOBE \\ 
\hline
\textbf{Count}  & 2285 & 1541 & 1616 & 1749 & 1788  \\
\end{tabular}
\caption{Number of occurrences for each label/region in dataset (8979 total count). This dataset is split into 60\% training data (5387 total count), 20\% validation data (1796 total count) and 20\% test data (1796 total count).}
\label{table:1}
\end{table}

Since we are identifying the main magnetospheric regions, a small dataset will perform well with deep learning models since the patterns are very similar for each region. The boundaries (magnetopause, bow shock and plasma sheet boundary layer) will be identified as a post processing step. If boundaries were included in the labels to be identified by the deep learning model, then a very large dataset would be necessary as the boundary location could be varying in each sample and the model would need many varying samples to be able to identify the patterns as this was the case by the model developed by Breuillard et al (2020) with 34,159 labels.

\subsection{Data Preprocessing}

\begin{figure}
\centering
\includegraphics[width=.99\textwidth]{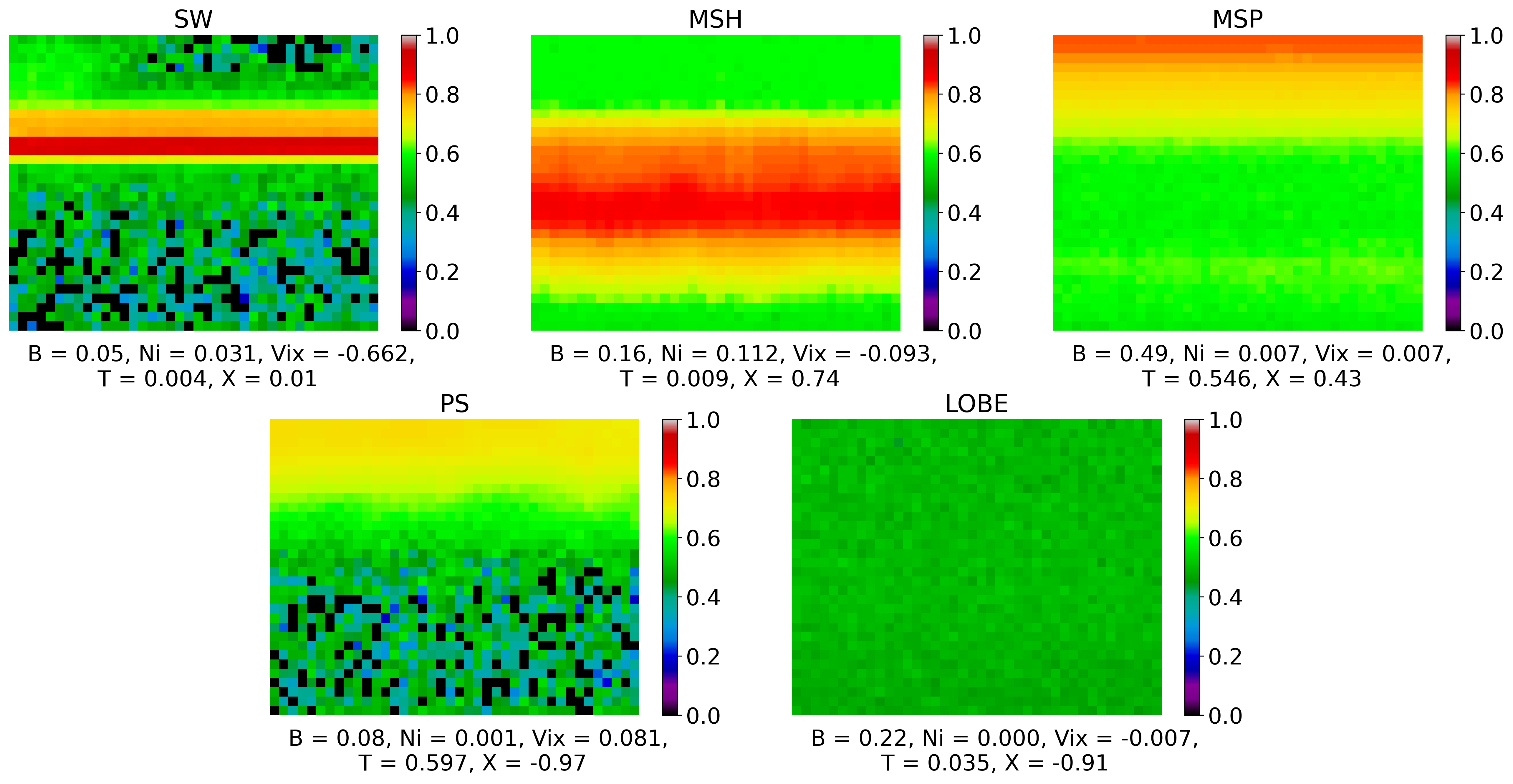}
\caption{\label{fig:regions} Magnetospheric regions corresponding to normalized ion energy spectrogram data and associated normalized scalar parameters.}
\end{figure}

Ion energy spectra possess three dimensions: time series (40 points), energy bins (32 bins), and flux magnitude. Plasma parameters (magnetic field, ion temperature, ion density, ion velocity component and position) are two-dimensional with time series and parameter values. All positive quantities (energy spectra, magnetic field magnitude, total ion temperature, ion density) undergo min-max normalization to values between 0 and 1. Ion velocity and position data are normalized between -1 and 1 using a piecewise linear transformation (since the original range is not symmetric around zero) to preserve directional information, for example distinguishing dayside (positive X) from nightside (negative X) locations or earthward flows (negative velocity) from tailward flows (positive velocity). An example for each label is shown in figure \ref{fig:regions} with normalized ion energy spectrogram data and associated plasma parameters. 

\section{Hybrid Model Design}

Machine learning projects are often restricted by the labor intensive process of collecting, cleaning, and labeling data, a critical step that directly impacts model performance and generalization. We show that our model needs minimal training data and it is computationally light. Our approach employs a novel hybrid architecture combining two complementary machine learning techniques optimized for different data characteristics. Ion energy spectrograms are processed using a Convolutional Neural Network (CNN), while scalar plasma parameters are analyzed using a Random Forest classifier. The CNN model identifies the energy patterns in ion energy spectrogram like a trained scientist would do. But the CNN model is not solely enough to identify all regions since regions like inner magnetosphere and plasma sheet could have similar patterns in ion energy spectrogram as shown in figure \ref{fig:regions}. These regions have different values in magnetic field, ion moments and their locations. That is why we use the random forest model on plasma parameters so we are able to clearly distinguish all the regions. A hybrid model like this will result in high accuracy while keeping the model lightweight.

\subsection{Convolutional Neural Network for Flux Data}

Ion energy spectrograms are treated as image-like data with dimensions (1, 32, 40) representing channels, energy bins, and timeseries (3-minute resolution), respectively. We employ a CNN architecture based on the Visual Geometry Group (VGG) design, characterized by deep structure utilizing small 3×3 convolutional filters \cite{vgg}. VGG models have been instrumental in advancing the field of image classification and have been used as a baseline for many subsequent models.  
The CNN processes the ion energy spectra data through multiple convolutional layers, enabling pattern recognition across both temporal and energy dimensions. This approach captures the characteristic spectral signatures that distinguish different plasma environments. Convolutional layers extract hierarchical features from simple edges and textures in early layers to complex patterns in deeper layers. Pooling layers provide dimensionality reduction while preserving essential spatial relationships. By learning and combining these features, CNNs can accurately recognize objects, faces, or other visual patterns within images. 

\begin{figure}
\centering
\includegraphics[width=.99\textwidth]{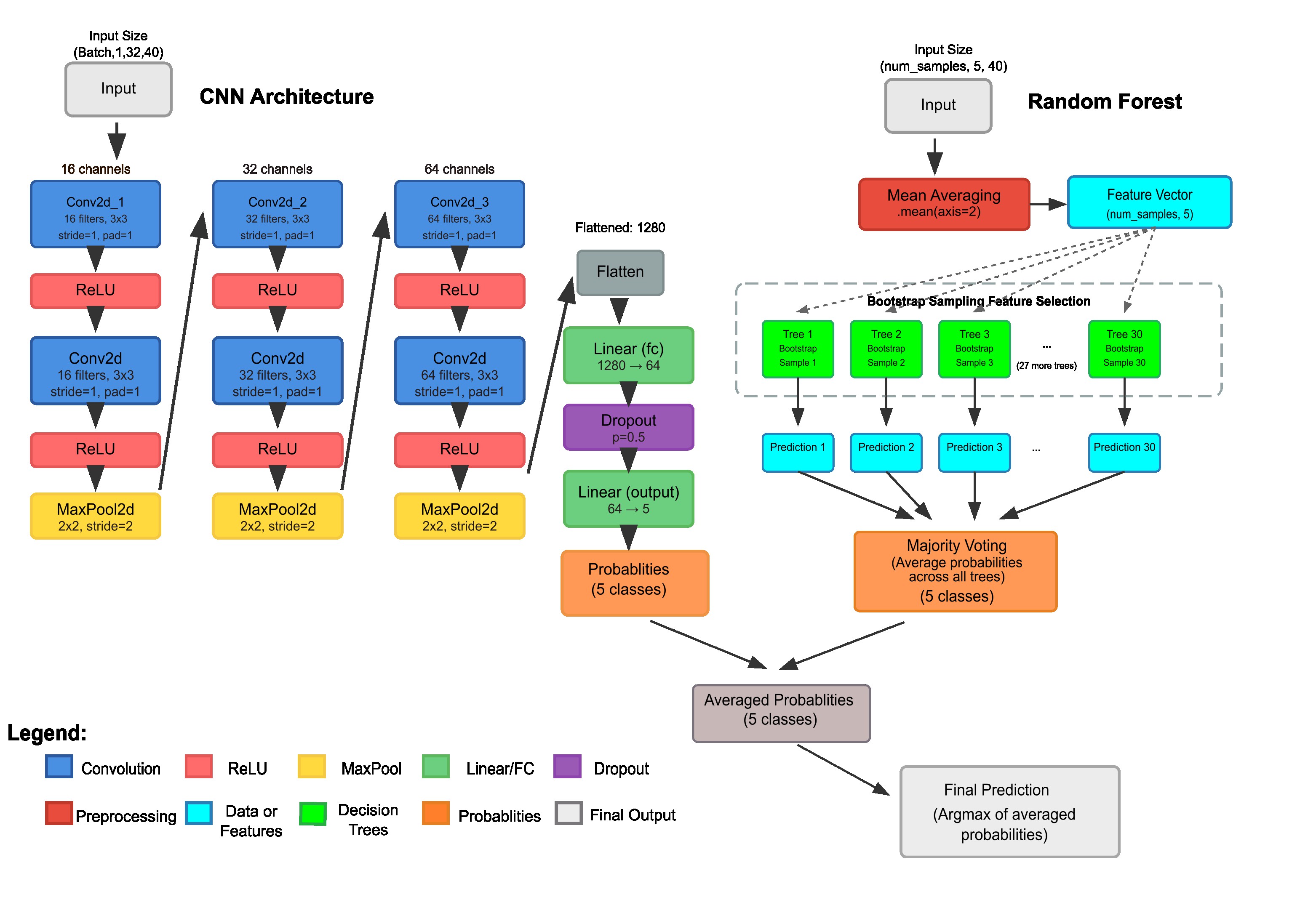}
\caption{\label{fig:cnnmodel} Summary of combined CNN and random forest models.}
\end{figure}

\begin{figure}
\centering
\includegraphics[width=.99\textwidth]{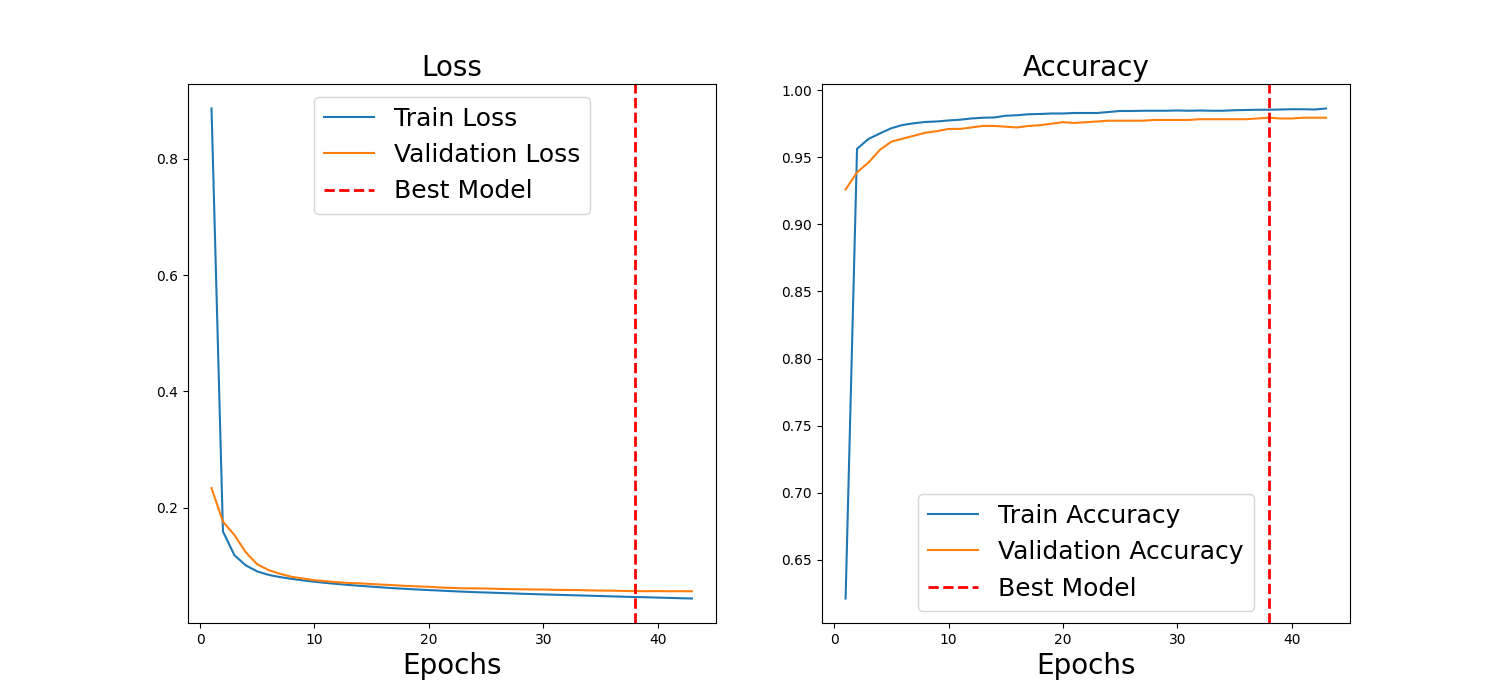}
\caption{\label{fig:loss} Summary of loss and accuracy for CNN model. The accuracy is 98\% on the the validation data.}
\end{figure}

Figure \ref{fig:cnnmodel} shows the details of CNN model applied to ion energy spectrogram data. The model includes three convolutional blocks with two convolutional layers each. It is designed for multi-class classification with 5 output labels. It uses Rectified Linear Unit (RELU) \cite{relu} activation function to add non-linearity and it randomly drops 50\% of neurons during training to prevent overfitting. We use optimizer Adam in our model \cite{adamoptimizer} and the learning rate is set to $5^{-5}$. This CNN model is trained for 43 steps as shown in figure \ref{fig:loss} using early stopping method. If there are no improvement in validation accuracy for 5 consecutive epochs, the model stops training and restore the best model weights before finishing. We achieve the accuracy of 98\% on the validation data in the CNN model at step 38.
The loss in figure \ref{fig:loss} is calculated using cross entropy function which is the difference between the predicted probability distribution (from CNN model) and the true distribution (the correct label). As figure \ref{fig:loss} shows both training loss (blue) and validation loss (orange) decrease quickly in the first few epochs which means model is learning effectively. After about epoch 5, both losses flatten near 0.1 or lower, indicating the model has almost converged. Overall, the model trains very effectively: sharp decrease in loss, sharp increase in accuracy and it reaches 98\% accuracy within 38 steps.

\subsection{Random Forest for Scalar Parameters}

Plasma parameters (magnetic field magnitude, ion denssity, ion velocity in X GSE, total ion temperature, position in X GSE) with dimensions (1, 40) are processed using a Random Forest classifier. The dimension 40 represents the 3-minute labeled interval with resolution of 4.5 seconds. These parameters are averaged in each labeled interval and are inputs into Random Forest model as scalar values. This ensemble method constructs 30 decision trees using bootstrap aggregation and random feature selection at each split. This randomness helps reduce overfitting and improves generalization. By aggregating the predictions of many diverse trees, random forests achieve higher accuracy and robustness compared to single decision trees. Figure \ref{fig:cnnmodel} shows the details of Random Forest model with the plasma parameters as input.

For each 3-minute interval, we compute average values of the three input parameters. The Random Forest model builds the decision trees based on these averaged quantities, capturing the characteristic parameter ranges associated with each magnetospheric region. This approach proves particularly effective for the discrete parameter boundaries that often distinguish plasma environments. This model achieves 98\% accuracy on the validation data.

\subsection{Model Integration and Prediction}

The final classification combines outputs from both models through probability averaging as shown in figure \ref{fig:cnnmodel}. Each component model generates probability distributions for the five region classes. These probabilities are simply summed and averaged to produce the final prediction, with the maximum probability determining the predicted region label.
This ensemble approach leverages the complementary strengths of both methods: the CNN's capability for complex pattern recognition in spectral data and the Random Forest's effectiveness in handling discrete parameter boundaries. The integration provides robust predictions that are less susceptible to individual model limitations.

\section{Results} 

\subsection{Model Performance Metrics}

\begin{figure}
\centering
\includegraphics[width=.99\textwidth]{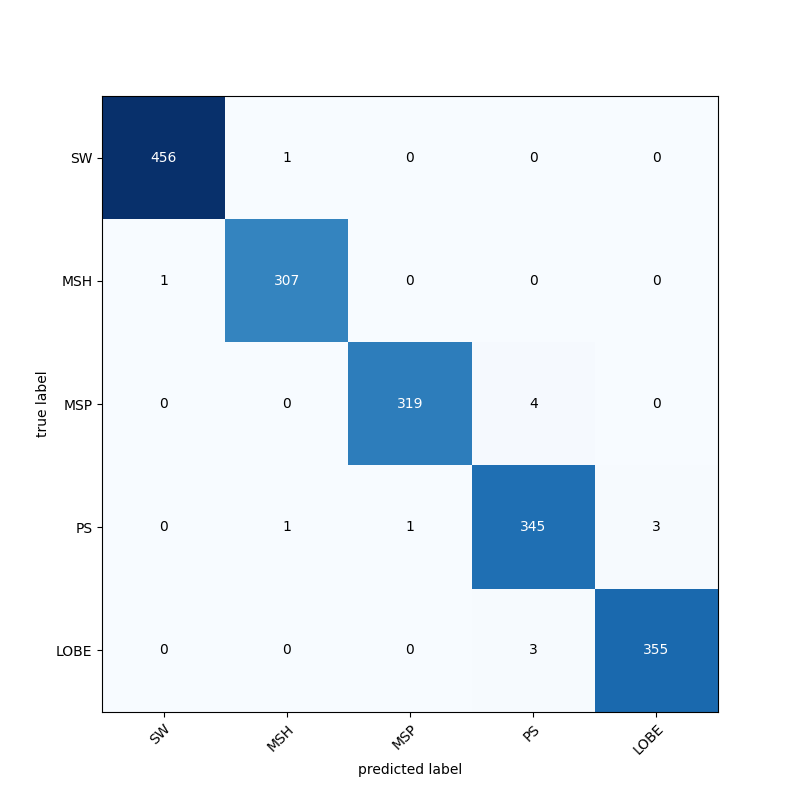}
\caption{\label{fig:conf} Confusion matrix applied to test dataset.}
\end{figure}

We evaluate model performance using standard machine learning metrics including accuracy, precision, recall, F1 score on a separate test data that the model has not seen. The confusion matrix provides detailed insights into classification performance across all region classes. A confusion matrix is a table used to evaluate the performance of a classification model by comparing its predicted labels to the actual labels. It organizes results into four categories: True Positives (TP), where the model correctly predicts a positive case; True Negatives (TN), where it correctly predicts a negative case; False Positives (FP), where it incorrectly predicts positive when the actual label is negative; and False Negatives (FN), where it predicts negative when the actual label is positive. Figure \ref{fig:conf} shows the confusion matrix on the test dataset indicating that the hybrid model has 14 incorrect prediction out of 1796 labels.

\begin{figure}
\centering
\includegraphics[width=.99\textwidth]{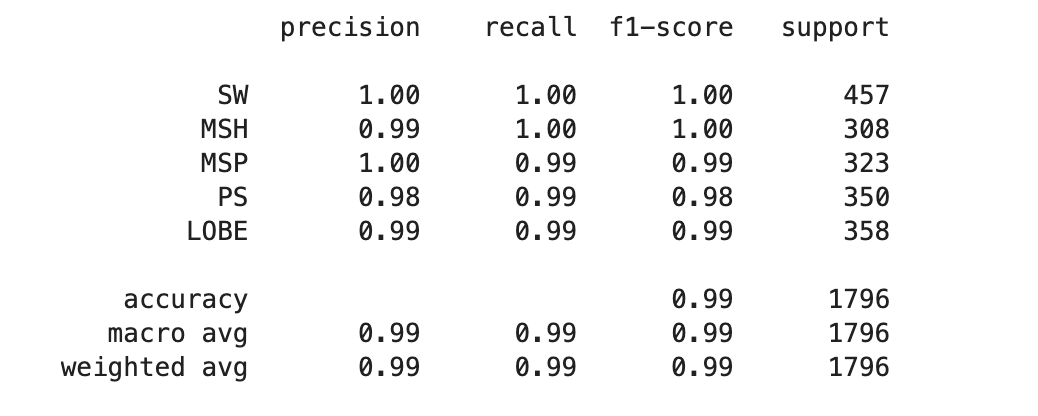}
\caption{\label{fig:auc} Classification report on test dataset. The hybrid model shows 99\% accuracy on the test dataset with an average F1 score of 0.99.}
\end{figure}

Accuracy (Correct predictions / Total predictions), precision (TP / (TP + FP)), recall (TP / (TP + FN)) and F1 Score (2 × Precision × Recall / (Precision + Recall)) are calculated using the test dataset as shown in figure \ref{fig:auc}. Our hybrid model achieves 99\% accuracy on the test dataset with an averaged F1 score of 0.99, demonstrating exceptional performance across all magnetospheric regions. The high F1 score indicates balanced precision and recall, suggesting the model effectively minimizes both false positive and false negative classifications. This performance significantly exceeds previous automated classification approaches while requiring substantially less training data. Table \ref{table:compare} shows a comparison of performance and comprehensiveness of current machine learning models to automatically identify magnetospheric regions using observational data.  

\begin{table}[h!]
\centering
 \begin{tabular}{||c c c c c||} 
 \hline
 Method & Inputs & Accuracy & Regions & Dataset \\ [0.5ex] 
 \hline\hline
 Breuillard 2020 (FCN) & 12 params & 89\% & 10 & large \\
 \hline
 Olshevsky 2021 (CNN) & 3D PSD & 98\% & 3 & large  \\
 \hline
 Nguyen 2022 (GBoost) & 8 params & 99\% & 3 & light \\
 \hline
 Toy-Edens 2024 (Clustering) & line spectra & 98\% & 4 & light\\
 \hline
 This work (CNN+RF) & spectra + 5 params & 99\% & 5 & light \\ [1ex] 
 \hline
 \end{tabular}
 \caption{\label{table:compare} Comparison of different machine learning models for automatic identification of magnetospheric regions.}
\end{table}

\subsection{Boundary Identification}

Beyond region classification, our approach enables automated boundary detection through identification of region transitions. Magnetopause (MP) crossings are identified through MSH $\leftrightarrow$ MSP transitions, bow shock (BS) crossings through MSH $\leftrightarrow$ SW transitions, and plasma sheet boundary layer (PSBL) crossings through PS $\leftrightarrow$ LOBE transitions.
Boundary intervals are assigned 6-minute duration to capture transitional characteristics, while main region intervals maintain 3-minute resolution. This temporal resolution can be easily adjusted based on specific analysis requirements or mission constraints.

\subsection{Operational Application}

We demonstrate model applicability across diverse orbital configurations including dayside, flank magnetopause and magnetotail regions. The model successfully identifies plasma environments throughout MMS orbital coverage, providing consistent performance across varying plasma conditions and geometric configurations. 

\begin{figure}
\centering
\includegraphics[width=.99\textwidth]{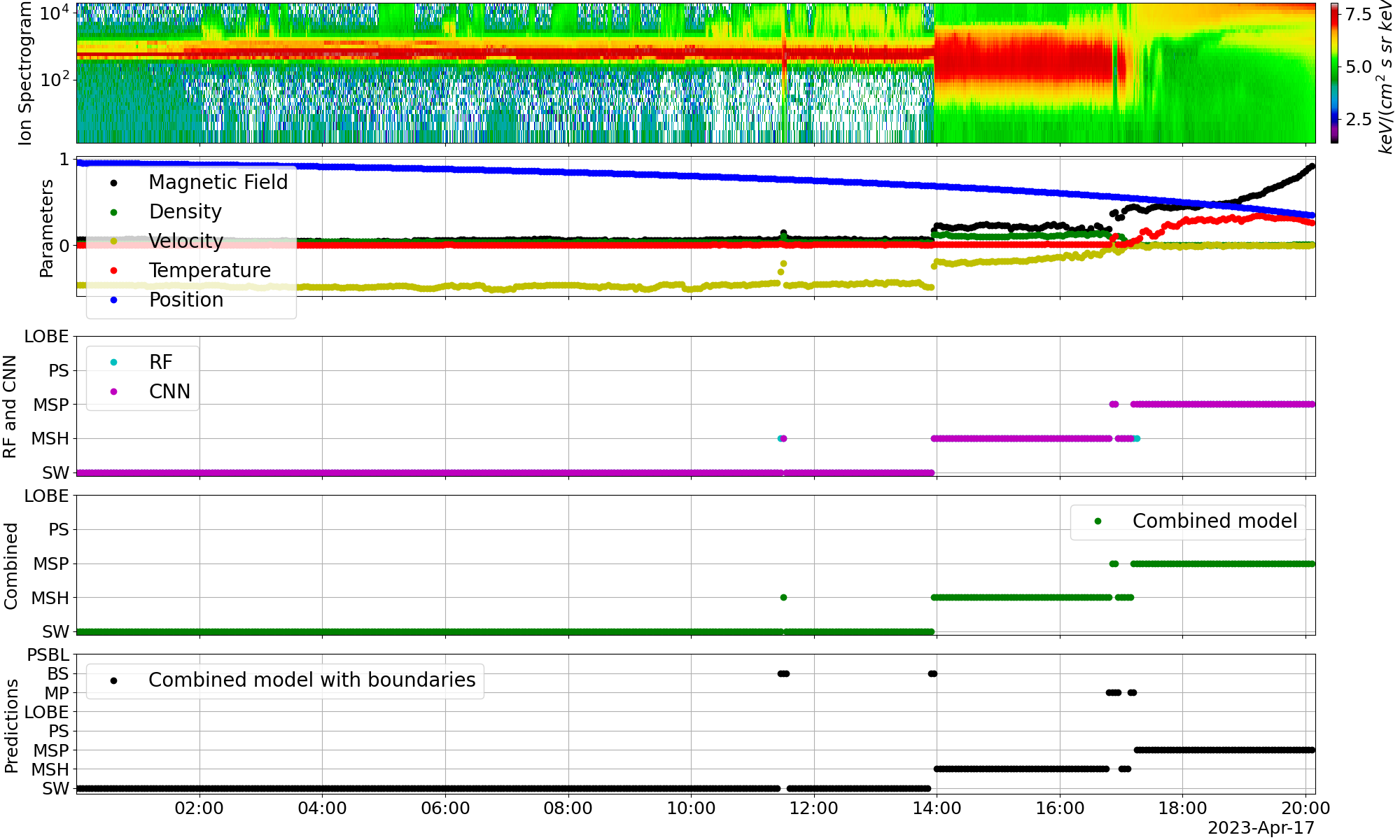}
\caption{\label{fig:day} Prediction results for dayside.}
\end{figure}

Figure \ref{fig:day} presents MMS1 observations from the dayside region on April 17, 2023. The panels display, from top to bottom: omnidirectional ion energy spectra, five normalized parameters (total magnetic field, ion density, ion velocity in X GSE, total ion temperature, and spacecraft position in X GSE), individual CNN and Random Forest model predictions, the combined prediction of the hybrid model and the final hybrid model classifications including the boundary transitions. Universal Time is shown on the bottom x-axis. The spacecraft trajectory demonstrates a complex sequence of boundary crossings. Initially positioned in solar wind, MMS1 encounters a brief bow shock crossing before returning to solar wind conditions. Subsequently, the spacecraft traverses the bow shock into the magnetosheath, followed by multiple magnetopause crossings before entering the inner magnetosphere. The hybrid model successfully identifies fine-scale boundary dynamics, including individual magnetopause and bow shock transitions.

\begin{figure}
\centering
\includegraphics[width=.99\textwidth]{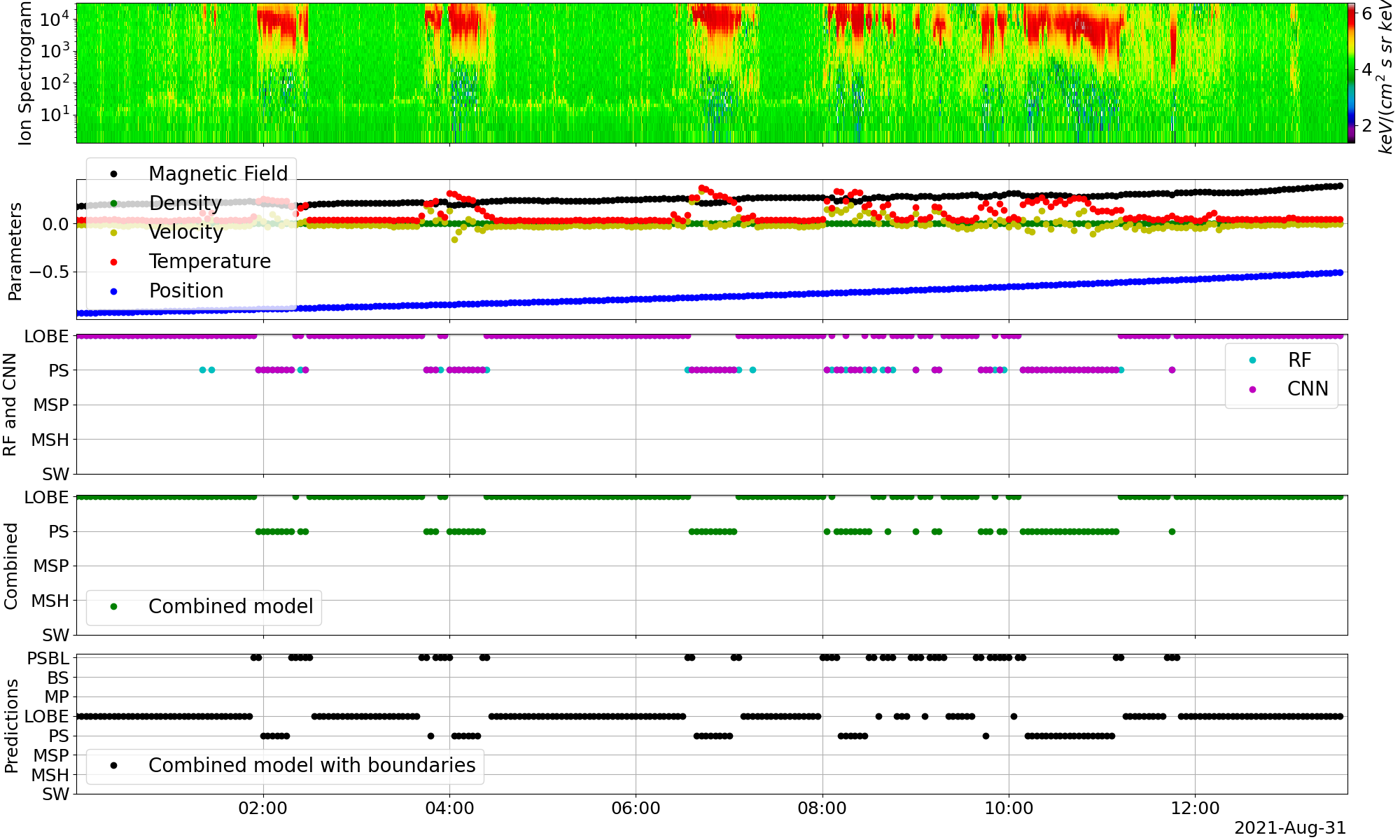}
\caption{\label{fig:night} Prediction results for nightside.}
\end{figure}

Figure \ref{fig:night} demonstrates the classification methodology for nightside observations, showing MMS1 data from the complete day of August 31, 2021, using the same panel format as Figure \ref{fig:day}. During this magnetotail pass, the hybrid model accurately identifies the distinct plasma environments encountered by MMS1, including lobe regions, many plasma sheet boundary layers, and plasma sheet, demonstrating the model's capability across different magnetospheric sectors.

\begin{figure}
\centering
\includegraphics[width=.99\textwidth]{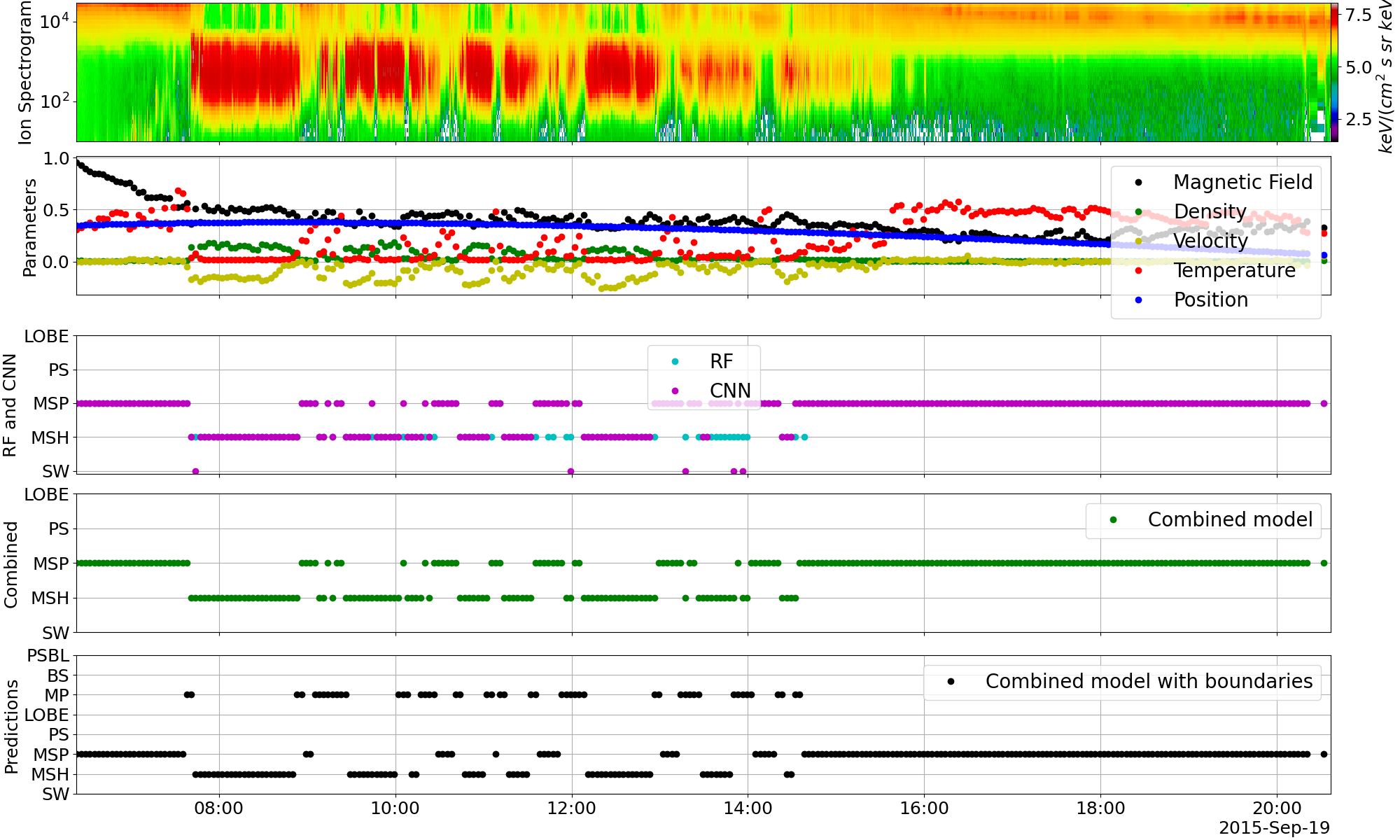}
\caption{\label{fig:flank} Prediction results for flanks.}
\end{figure}

Figure \ref{fig:flank} demonstrates the classification methodology for flank magnetopause observations with many boundary crossings, showing MMS1 data from the complete day of September 19, 2015, using the same panel format as Figure \ref{fig:day}. During this flank magnetopause pass, the hybrid model accurately identifies the plasma environments encountered by MMS1, including inner magnetosphere regions, magnetosheath, and many frequent magnetopause crossings, showing the model's capability to perform in complex scenarios. In this case, the CNN model misclassified some intervals as solar wind but random forest model had the correct predictions which helps the combined model to correctly identify the regions. 

These three case studies corroborate the quantitative performance metrics by demonstrating the hybrid model's ability to accurately classify the majority of plasma regions encountered across MMS orbital trajectories. The examples validate the model's practical applicability to operational data analysis workflows.

\section{Discussion and Future Directions}

We present a lightweight supervised machine learning approach for automated identification of magnetospheric regions using MMS mission data. Our hybrid model combines convolutional neural networks for ion energy spectra analysis with Random Forest classification for scalar plasma parameters, achieving 99\% accuracy with minimal training data requirements.

The approach successfully identifies five primary magnetospheric regions (solar wind, magnetosheath, magnetosphere, plasma sheet, and lobe) at 3-minute temporal resolution. Beyond region classification, we detect the boundaries (magnetopause, bow shock and plasma sheet boundary layers) through identification of region transitions.
Boundary intervals are assigned 6-minute duration combining two main region intervals. This temporal resolution can be adjusted based on specific analysis requirements or mission constraints. The computational efficiency and high accuracy make this method suitable for both scientific analysis and operational applications. 

Our work presents several significant advances over existing magnetospheric region classification methods. While Breuillard et al. (2020) achieved 89\% accuracy using a fully convolutional network with 12 plasma parameters and 34,159 training samples, our hybrid CNN-Random Forest architecture achieves 99\% accuracy with only 8979 training samples by strategically combining spectrogram pattern recognition with scalar parameter analysis. Unlike domain specific approaches such as Olshevsky et al. (2021) and Wang et al. (2025), which focus exclusively on dayside regions, our method provides comprehensive coverage across all five primary magnetospheric regions spanning both dayside and nightside environments with consistent performance. Furthermore, our methodology uniquely enables automated boundary detection (magnetopause, bow shock, plasma sheet boundary layer) as an inherent capability through region transition identification, eliminating the need for separate boundary detection models as employed in prior studies \cite{Breuillard2020, Argall2020}.

While our model demonstrates excellent performance on MMS data, generalization to other missions requires validation and potential retraining. The relatively small training dataset, while sufficient for current performance, may benefit from expansion to improve robustness across extreme and varying conditions.
Future developments may extend the approach to include additional plasma parameters, higher temporal resolution analysis, or multi-spacecraft coordinated observations. Integration with physics based models could provide enhanced understanding of the physical processes governing region boundaries and transitions.

This work demonstrates the potential for machine learning approaches to automate fundamental data processing tasks in space physics, enabling more efficient analysis of the growing volumes of observational data from current and future magnetospheric missions.

\section*{Data Availability}

All the MMS data used in this work are available at the MMS data center at \url{https://lasp.colorado.edu/mms/sdc/public/about/browse-wrapper/}. The data have been loaded, analyzed, and plotted using the PYSPEDAS software, which can be downloaded via the Downloads and Installation page (\url{https://github.com/spedas/pyspedas}).

We used PyTorch and scikit-learn packages in developing our machine learning models. PyTorch is used for developing the CNN model and the scikit-learn is used for Random Forest model. Complete model implementation and example applications are available in the GitHub repository (\url{https://github.com/nargesahmadi/Magnetospheric-Regions}).

\section*{Acknowledgments}

This work was supported by NASA MMS mission funding. The authors thank the MMS instrument teams for providing high-quality data products and the PYSPEDAS team for data analysis tools. Figures were produced using matplotlib \cite{Hunter2007}. Machine learning models were implemented using PyTorch 2.2 and scikit-learn.


\end{document}